\documentclass[letterpaper,aps,prl,twocolumn,amsmath,showpacs,amssymb]{revtex4}

\usepackage{color}
\usepackage[pdftex]{graphicx,hyperref}% Include figure files
\usepackage{dcolumn}% Align table columns on decimal point
\usepackage{bm}% bold math
\usepackage{marvosym}
\usepackage{times,amsmath,amssymb}
\begin{document}

\title{ATR excitation of surface polaritons at the interface between a metal and a layer of nanocrystal quantum dots}

\author{Yu. V. Bludov, M. I. Vasilevskiy}

\affiliation{Centro de F\'{\i}sica e Departamento de F\'{\i}sica, Universidade do Minho,
Campus de Gualtar, Braga 4710-057, Portugal}
\begin{abstract}
Surface plasmon-polaritons in a multilayer structure consisting of a metallic film and one or more layers of nanocrystal (NC) quantum dots (QDs) are studied. It is shown that there is resonance coupling between the plasmon-polaritons propagating along the metal/NC-layer interface and excitons confined in the dots, which makes a considerable effect on the optical properties of the structure unless the dispersion of the QD size is too large. This coupling can be explored in order to selectively excite QDs of different size by using an attenuated total reflection (ATR) structure. It opens the possibility of control of the relative intensity of light of different color, emitted by the QDs of different size.

\end{abstract}
\pacs{73.20.Mf, 71.36.+c, 78.67.Bf}
\maketitle

\section{Introduction}

Colloidal chemistry techniques have proved the ability to sinthesize high quality nanocrystals (NCs) of II-VI semiconductors, possesing the properties of quantum dots (QDs). With these NCs, a broad variety of nanostructures can be prepared, including multilayer structures of QDs of different average size, deposited on different substrates \cite {Rogach2008}. Combining such structures with other materials, such as organic dielectrics \cite {Basko2000} or graphene \cite{Chen2010}, can result in new interesting physics and applications related to the energy transfer from photoexcited NCs to these materials.
Recent experiments \cite {Gomez10} demostrated the existence of a strong coupling between excitons confined in nanocrystal quantum dots (QDs) and surface plasmon-polaritons (SPPs) propagating along the interface of a silver film and the QD layer deposited on top of it. The observation was achieved by measuring attenuated total reflection (ATR) spectra of the structure. In this work we develop the theory of the observed effect taking into account the the QD size dispersion. We will show that the resonance SPP coupling excitons confined in QDs can be considerable unless the dispersion of the QD size is too large or the dots are too far from the metal/dielectric interface. The resoanance coupling can be used for controllable pumping the dots in order to explore their uncomparable luminescence properties \cite {Rogach2008}. Based on the calculated results, we will discuss possible applications of ATR structures containing, beyond the silver film, several layers of NCs of different average size.

\section{Dielectric function of a composite material containing quantum dots}

A single QD can be described by the electronic susceptibility taking into account inter-band transitions:
\begin{equation}
\label{chi}
\chi (\omega )=\frac {4 \left \vert d_{cv} \right \vert ^2} {3V} \sum _n {\left \vert \beta_{n}
\right \vert ^2} \left [ \frac 1 {E_n-i\Gamma-\hbar \omega } + \frac 1 {E_n-i\Gamma+\hbar \omega } \right ] ,
\end{equation}
where the sum runs over confined exciton states with energies $E_i$ ($i=0$ denotes exciton vacuum), $d_{cv}$ is the transition dipole moment matrix element between valence and conduction bands of the underlying semiconductor material, $V$ is the QD volume and
$$
\beta _i=g_i\int_{QD} \Psi_{ex}^{(i)}\left (\mathbf r,\mathbf r\right ) d\mathbf r
$$
with $g_i$ and $\Psi_{ex}^(i)\left (\mathbf r_e,\mathbf r_h\right )$ denoting the degeneracy factor and the wavefunction of the corresponding exciton states.

The quantity
\begin{equation}
\label{epsilon}
\epsilon _{QD}(\omega )=\epsilon _s^{\infty}+4\pi \chi (\omega )
\end{equation}
can be regarded as a QD dielectric function, where $\epsilon _s^{\infty}$ is the high-frequency dielectric constant of the QD material. However, it can be more convenient to define the QD polarizability (assuming its spherical shape),
\begin{equation}
\label{alpha}
\alpha (\omega )=\frac {\epsilon _{QD} (\omega )-\epsilon _h}{\epsilon _{QD} (\omega )+2\epsilon _h}a^3\:,
\end{equation}
where $a$ is the QD radius and $\epsilon _h$ is the dielectric constant of the host material. Using these quantities (\ref{epsilon},\ref{alpha}), one can calculate an effective dielectric function, $\epsilon ^\star $, of the composite material containing uniform size QDs embedded in the host matrix using one of the following schemes:

(i) Maxwell-Garnett approximation (MGA) valid in the low QD concentration limit \cite {MG},
\begin{equation}
\label{MG}
\frac {\epsilon ^\star -\epsilon _h}{\epsilon ^\star+2\epsilon _h}=\frac {4\pi } 3 N\alpha \: ;
\end{equation}

(ii) Bruggemann mean field approximation (BA) \cite {Bruggeman},
\begin{equation}
\label{BA}
f\frac {\epsilon ^\star -\epsilon _{QD}}{\epsilon ^\star+2\epsilon _{QD}}+(1-f)\frac {\epsilon ^\star -\epsilon _h}{\epsilon ^\star+2\epsilon _h}
=0\: ,
\end{equation}
where $f=\frac {4\pi } 3 Na^3$ is the volume fraction of QDs. BA is valid when $f\sim 0.5$. MGA can be extended up to $f \sim 0.1$ using a renormalized polarizability which takes into account dipole-dipole interactions between QDs as explained in Ref. \onlinecite {VA96}.

The modified MGA (MMGA) proposed in in Refs. \onlinecite {VA96,MV00} allows for taking into account the QD radius dispersion. Let $F(a)$ denote the radius distribution function. The renormalized polarizability, $\alpha ^\star $, can be calculated by the following equations,
\begin{equation}
\label{alphastar}
\alpha ^\star =\int {\alpha (a^\prime)\frac {1-\sqrt { 1-4\Theta(a^\prime)}}{2\Theta(a^\prime)}}F(a^\prime)da^\prime \: ;
\end{equation}
\begin{equation}
\label{Theta}
\Theta (a^\prime)=\frac {8\pi}3 N\alpha (a^\prime)\int {\frac {\alpha (a)}{(a^\prime+a)^3}}F(a)da
\end{equation}
where $\alpha (a)$ is the polarizability of a QD of radius $a$. For uniform-size QDs Eqs. (\ref {alphastar}) and (\ref {Theta}) reduce to
$$
\alpha ^\star =\frac {2a^3}{f\tilde \alpha }\left (1-\sqrt { 1-f\tilde \alpha ^2 }\right )
$$
where $\tilde \alpha =\alpha /a^3$. $\alpha ^\star $ replaces $\alpha $ in Eq. (\ref {MG}) [notice that for small QD volume fraction $f$ $\alpha ^\star=\alpha$].

\section{The case of CdSe spherical QDs}

Within the simplest model neglecting the Coulomb interaction between the electron and hole (strong confinement limit) and multiple sub-band structure of the valence band, the QD exciton spectrum is given by \cite {Brus},
\begin{equation}
\label{exciton}
E_i=E_g+\frac {\hbar ^2 \xi _i ^2}{2 \mu a^2}
\end{equation}
where $E_g$ is the band gap energy (=1.75 eV for CdSe), $\mu $ is the electron-hole reduced mass and $\xi _1=\pi$, $\xi _2\approx 4.49$, $\xi _3\approx 5.76,\dots $ are the roots of the spherical Bessel functions. At the same time, $\beta _i=2$ for all $i\geq 1$. The dipole moment matrix element can be expressed as
$$
d_{cv}=\frac {e\hbar}{im_0E_g}p_{cv}
$$
where $m_0$ is the free electron mass and $2p_{cv}^2/m_0\approx $20 eV. We take $\epsilon _h=1.5$ for PMMA matrix and assume a Gaussian distribution of QD radius,
$$
F(a)=\frac 1 {\sqrt {2\pi \Delta _a^2}}\exp{\left [-\frac {(a-\bar a)^2}{2 \Delta _a^2}\right ]}
$$
with $\bar a$ and $\Delta_a$ being the average radius and the dispersion, respectively. The real and imaginary parts of the effective dielectric function of the QD/PMMA composite calculated for $f=0.1$ is shown in Fig. 1. A small imaginary part, $\Gamma =1$ meV was introduced in the denominators of Eq. (\ref {chi}).

\begin{figure}
\begin{center}
\includegraphics*[width=8cm]
{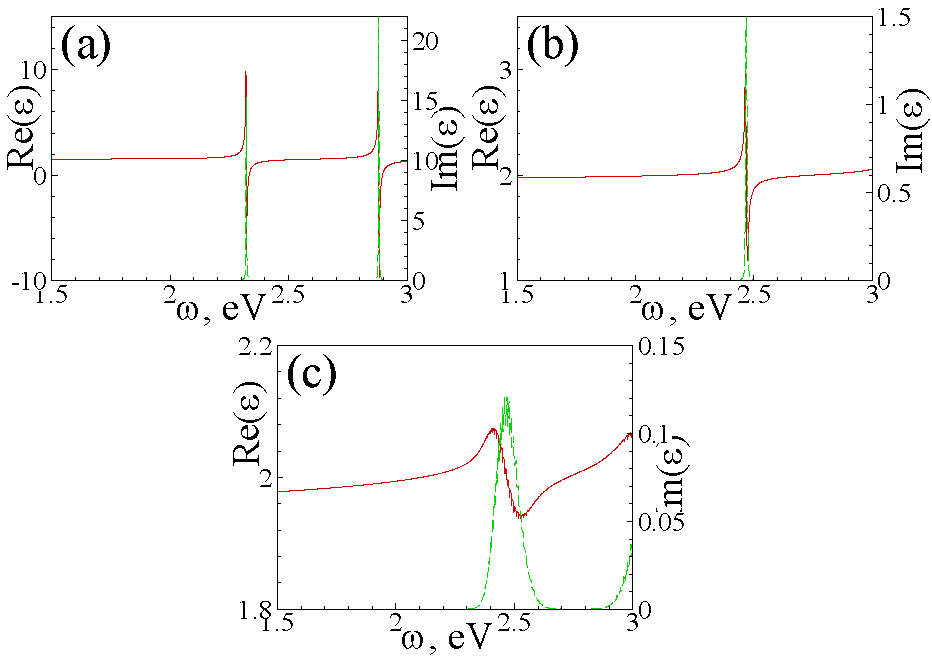}
\end{center}
\caption{ (color on line) Real (solid lines) and imaginary (dashed lines) parts of the CdSe QD - PMMA composite dielectric function for average QD radius $\bar a=3\,$nm and filling factor $f=0.1$, calculated using MMGA with QD radius dispersion $\Delta _a=0$ (panel a), $\Delta _a=0.005\bar a$ (panel b), and $\Delta _a=0.1\bar a$ (panel c).
} \label{fig:eps}
\end{figure}

As it can be seen from Fig. 1, the imaginary part of the effective dielectric function has a resonance at the exciton frequency, with the broadening depending on the QD concentration and size dispersion. The real part of $\epsilon ^\star$ can reach negative values in the vicinity of the exciton frequency only for vanishing dispersion [Fig. \ref{fig:eps}(a)], while for larger dispersion it always remains positive [compare with Figs. \ref{fig:eps}(b) and \ref{fig:eps}(c)].

\section{SPP at the metal/NC-layer interface}

\begin{figure}
\begin{center}
\includegraphics*[width=8cm]
{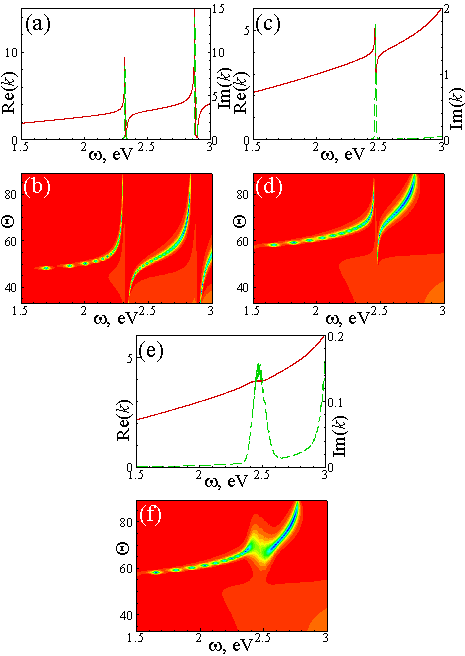}
\end{center}
\caption{(color on line) Panels {\bf (a, c, e):} Real (solid lines) and imaginary (dashed lines) parts of the SPP wavevector versus frequency, calculated for the interface between semiinfinite CdSe QD - PMMA composite and silver; Panels {\bf (b, d, f):} reflectivity $R$ \textit{versus} angle of
incidence $\theta$, and frequency $\omega$ for the ATR structure with a glass prism ($\varepsilon_g=2.9584$), a silver film of thickness $d=53.3\,$nm, and a semiinfinite composite medium. The composite dielectric function was calculated using MMGA with $\bar a=3\,$nm and $\Delta _a=0$ (panels a, b), $\Delta _a=0.005\bar a$ (panels c, d), and $\Delta _a=0.1\bar a$ (panels e, f).}
\label{fig:reflectance}
\end{figure}

In order to demonstrate the involved physics, we shall first derive the dispersion relation for surface plasmon-polaritons \cite{c:SPP-theory} in a simplified structure. Let us consider an electromagneic wave in the vicinity of a plane interface between two semi-infinite media, a metal and a QD/dielectric composite, and assume that the wave is $p-$polarized. We choose $x$ axis along  the direction of propagation of the electromagnetic wave within the interface plane and $z$ axis perpendicular to the interface. Assuming the electromagnetic field's time- and $x$-coordinate dependence to be $\sim \exp{(ikx-i\omega t)}$ (where $k$ and $\omega$ are the wavenumber and frequency), we can write down the wave equation in the form:
\begin{equation}
\nonumber
\frac{\partial^2 E^{(m)}_x}{\partial z^2}-\left(k^2-\frac{\omega^2}{c^2}\varepsilon_m\right) E^{(m)}_x=0\:.
\end{equation}
It possesses solution corresponding to an evanescent wave,
\begin{eqnarray}
E^{(1)}_x(z)=E^{(1)}_x(0)\exp(-p_1 z), \label{eq:sol1}\\
E^{(2)}_z(z)=-\frac{ik}{p_2}E^{(2)}_x(0)\exp(p_2 z),
\label{eq:sol6}
\end{eqnarray}
where $p_m=\sqrt{k^2-\omega^2\varepsilon_m/c^2};\:m=1,2$ and ${\rm Re}(p_m)>0$ as it is characteristic of surface waves. Let $m=1$ stand for the metal and  $\varepsilon_1 \equiv \epsilon_M(\omega)$ is the Drude-type dielectric function, and $m=2$ corresponds to the composite medium, $\varepsilon_2 \equiv \epsilon ^\star(\omega)$.
By applying boundary conditions at the interface we obtain the SPP dispersion relation in the form:
\begin{equation}
\frac{\varepsilon_2}{p_2}+\frac{\varepsilon_1}{p_1}=0.
\label{eq:dr}
\end{equation}
It can be solved for the wavenumber $k$, yielding
$$
k=\frac{\omega}{c}\left(\frac{\varepsilon^*(\omega)\varepsilon_{M}(\omega)}{\varepsilon^*(\omega)+\varepsilon_{M}(\omega)}\right)^{1/2}\:.
$$
The frequency dependence of the real and imaginary parts of $k$ for an Au/QD--PMMA interface is shown in Figs. \ref{fig:reflectance}(a), (c), and (e).
${\rm Im} (k)$ increases drastically in the vicinity of resonance frequencies that are determined by the relation ${\rm Re} \left( \varepsilon^*(\omega)+\varepsilon_{M}(\omega)\right )=0$ and are quite close to the QD exciton transition frequencies. It reflects the resonant coupling between SPPs and QD excitons. Again, the resonance peak broadens with the increase of the QD size dispersion because of the weaker coupling involving fewer QDs for each given $\omega $.

\begin{figure}
\begin{center}
\includegraphics*[width=8cm]
{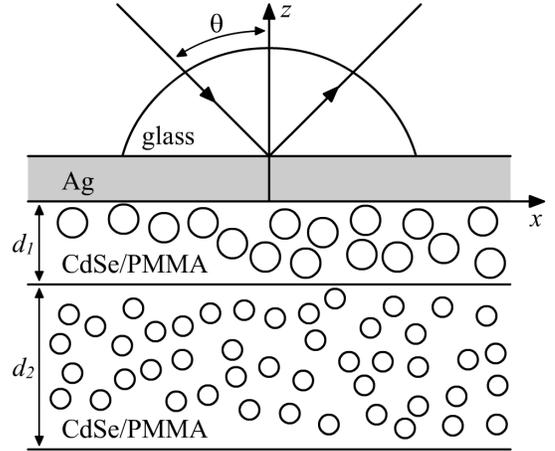}
\end{center}
\caption{ ATR structure composed of a glass prism, a silver plate, and two composite layers containing semiconductor NCs of different average size}
\label{fig:ATR}
\end{figure}

As known, SPPs can be probed in the ATR geometry schematically shown in in Fig. \ref{fig:ATR}, where the frquency and the $x$-projection of the wavevector of an external electromagnetic wave can match $\omega $ and ${\rm Re} (k)$ of the surface polaritons \cite{c:SPP-theory},
\begin{equation}
\nonumber
k_x=\frac \omega c \sqrt {\epsilon _g}\sin \theta\:,
\label{scanline}
\end{equation}
where $\epsilon _g $ is the dielectric constant of the prism and $\theta $ is tha angle of incidence. The matching is achieved by adjusting $\omega $ and/or $\theta $ and is detected by measuring the reflectance, $R$, which shows characteristic dips corresponding to the resonance transfer of the electromgnetic energy into surface polaritons. We have performed calculations of $R$ for the ATR structure consisting of a prism, an Ag film \footnote {For the dielectric function of silver we used the expression and parameters given in Ref. \onlinecite {Ag_df}},
and one or two layers of CdSe QD--PMMA composites (Fig. \ref{fig:ATR}), which are similar to those outlined above for the simplified structure containing just two semi-infinite media. The minima in the ATR spectra can be observed in Figs. \ref{fig:reflectance}(b),(d), and (f). They resemble the structure of the corresponding ${\rm Re}(k)$ \textit{vs} $\omega$ dependencies [Figs. \ref{fig:reflectance}(a), (c), and (e)].

\section{Results and discussion}

\begin{figure}
\begin{center}
\includegraphics*[width=8cm]
{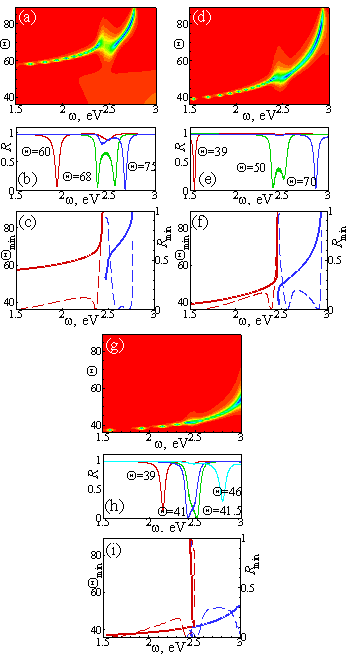}
\end{center}
\caption{(color on line) Reflectivity \textit{versus} angle of
incidence $\theta$ and frequency $\omega$ [panels {\bf (a, d, g)}]; reflectivity \textit{versus} frequency $\omega$ for a fixed $\theta$ [panels {\bf (b, e, h)}]; minimal reflectivity $R_{min}$ (dashed lines) and corresponding angle of incidence $\theta_{min}$ (solid lines) for a given $\omega$ [panels {\bf (c, f, i)}] for the ATR structure with glass prism $\varepsilon_g=2.9584$, silver film of thickness $d=53.3\,$nm, and CdSe QD--PMMA composite layer of different thickness: semiinfinite (a--c), $d_1=53.3\,$nm (d--f) and $d_1=19.78\,$nm (g--i). In the case of finite $d_1$, the medium at $z<-d_1$ is vacuum. The composite dielectric function was calculated using MMGA with $f=0.1$, $\bar a=3\,$nm and $\Delta _a=0.1\bar a$.}
\label{fig:reflectance-ft}
\end{figure}

Let us consider the structure of the SPP-exciton resonance in more detail. In the case of semiinfinite composite with CdSe NCs [Figs. \ref{fig:reflectance-ft}(a)--(c)] the resonant interaction of modes occurs at $\Theta\approx 68^o$ for the chosen set of parameters. In the vicinity of this angle the dependence $R(\omega)$ exhibits {\it two} minima with approximately equal depth, while for other angles of incidence one of the minima is significantly deeper than the other [see Fig. \ref{fig:reflectance-ft}(b)]. This is characteristic of mode anticrossing and corresponds to the experimental observation of Ref. \onlinecite {Gomez10}. The positions of these minima (i.e., the angles of incidence, $\theta_{min}$, at which the reflectance reaches its minimum, $R_{min}$, for a given frequency) are depicted in Fig. \ref{fig:reflectance}(c), where one can clearly see the resonant SPP-exciton interaction, or mode anticrossing. At the same time the depth of the resonant minima of $R_{min}(\omega)$ at the corresponding angles $\theta_{min}$ demonstrates that there are at least three points ($\omega\approx 2.36\,$eV, $\omega\approx 2.6\,$eV, $\omega\approx 2.72\,$eV) where the reflectivity of the ATR structure reaches zero, $R_{min}=0$.

How does the CdSe QD--PMMA layer thickness influence the ATR spectrum? First, the ATR resonance minima are shifted to lower angles of incidence [compare Figs. \ref{fig:reflectance-ft}(a) and (d)]. Secondly, the anticrossing of modes becomes weaker as manifested by the smaller separation of the reflectance minima in Fig. \ref{fig:reflectance-ft}(e) for $\theta\approx 50^o$, as well as in Fig. \ref{fig:reflectance-ft}(f)]. Thirdly, the values of $R_{min}$ far from the exciton resonance frequency are higher than in the case of semi-infinite composite medium. Further decrease of the CdSe QD--PMMA layer thickness [Figs. \ref{fig:reflectance-ft}(g)--(i)] leads to the damping of the SPP-exciton interaction. For example, in the case of Fig. \ref{fig:reflectance-ft}(g) the resonance is hardly distinguishable (it is confirmed also by Fig. \ref{fig:reflectance-ft}(i) where the positions of resonant minima are shown explicitly). Also from Fig. \ref{fig:reflectance-ft}(h) one can see that for the angles $\theta\approx 41^o$ and $\theta\approx 41.5^o$ two reflectivity minima practically merge and cannot be resolved.

\begin{figure}
\begin{center}
\includegraphics*[width=8cm]{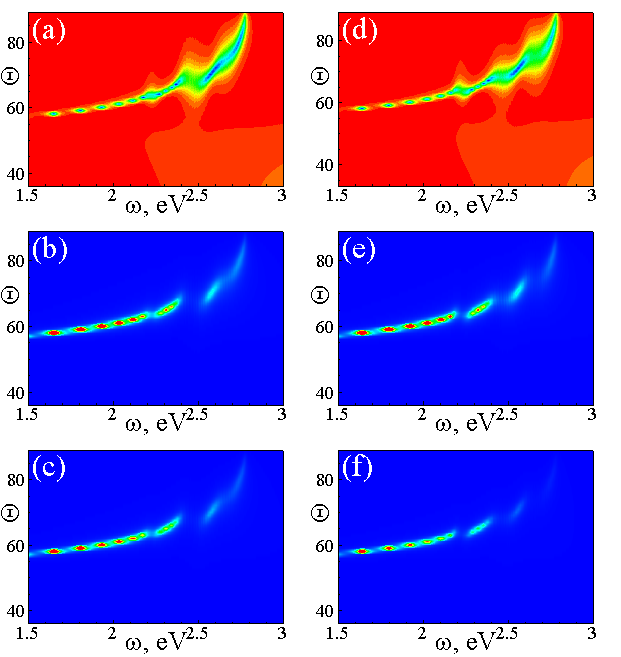}
\end{center}
\caption{(color on line) Reflectivity (a, d), and relative electric field square amplitude $|E_x(z)/E_i|^2$ at $z=-d$ (b, e) and $z=-(d+d_1)$ (c, f) \textit{versus} angle of
incidence and frequency for the ATR structure consisting of glass prism $\varepsilon_g=2.9584$, Ag film of thickness $d=53.3\,$nm and two CdSe QD--PMMA composite layers, one of thickness $d_1=19.78\,$ nm (a--c) or $d_1=53.3\,$nm (d--e) and average QD radius $a_1=3.7\,$nm, and the other semi-infinite, with average QD radius $a_2=3\,$nm. The dielectric functions for both composite layers were calculated using MMGA with $f=0.1$ and QD radius dispersion $\Delta_a=0.1\bar a$ in all cases.}
\label{fig:reflectance-2l}
\end{figure}

Now let us turn to the structures with the same silver film and {\it two} composite layers, one of thickness $d_1$ containing QDs of average radius $a_1$ and the other (semi-infinite) with QDs of radius $a_2$ (Fig. \ref{fig:ATR} with $d_2\rightarrow \infty$). In the ATR spectrum of this structure [Figs. \ref{fig:reflectance-2l}(a), (d)], one can observe resonances corresponding to both $a_1$ [$\hbar\omega\approx 2.2\,$eV and $\hbar\omega\approx 2.7\,$eV in Figs. \ref{fig:reflectance-2l}(a), (d)] and $a_2$ QDs [$\hbar \omega\approx 2.5\,$eV in Figs. \ref{fig:reflectance-2l}(a), (d)]. Increasing $d_1$ results in a stronger anticrossing between the SPP and the $a_1$ QD exciton modes, and in a weakeing of the $a_2$ resonance [compare Figs. \ref{fig:reflectance-2l}(a) and (d)]. Nevertheless, choosing an appropriate composite layer thickness, one can achieve SPP coupling to QD excitons localized in the composite layer which is not adjacent to the metallic film.

ATR minima correspond to some particular values of $\omega $ and $\theta $ for which the energy of the incident electromagnetic wave is most efficiently transferred to SPP and, consequently, to the QD excitons.
This is shown in Figs. \ref{fig:reflectance-2l}(b),(c),(e) and (f), where the squared average amplitude of the electric field is plotted for each of the CdSe QD--PMMA layers. It represents the intensity of excitation of QD luminescence, whose characteristic colour is determined by the QD size. Therefore one can adjust the relative PL intensity of different colours emitting by two or more composite layers containing QDs of different average radii. It opens the possibility of building a lightning device with colour characteristics which can be controlled either by the excitation frequency or by the ATR incidence angle.

\begin{figure}
\begin{center}
\includegraphics*[width=8cm]{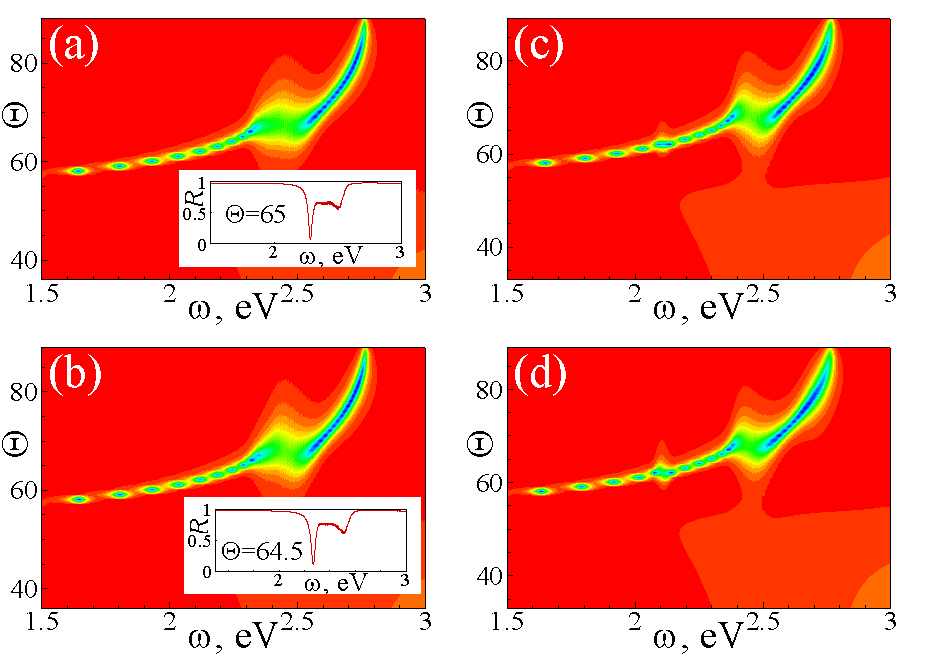}
\end{center}
\caption{(color on line) Reflectivity \textit{versus} angle of
incidence and frequency for the ATR structure consisting of glass prism $\varepsilon_g=2.9584$, Au film of thickness $d=53.3\,$nm and two CdSe QD--PMMA composite layers, one of thickness $d_1$ and average QD radius $a_1$ $bar a_1=4.2876\,$nm (c, d) and the other semiinfinite with average QD radius $\bar a_2=3\,$nm. The following parameters have been used: $d_1=53.3\,$nm, $a_1=3\,$nm $a_2=3.2\,$nm (a); $d_1=79.12\,$nm, $a_1=3\,$nm $a_2=3.2\,$nm (b); $d_1=19.78\,$nm, $a_1=4.2876\,$nm $a_2=3\,$nm (c); $d_1=53.3\,$nm, $a_1=4.2876\,$nm $a_2=3\,$nm (d). Insets in panels (a, b) present the dependence of the reflectivity on frequency for a fixed angle of incidence.}
\label{fig:reflectance-2l-eq}
\end{figure}

The average sizes of the NCs embedded in different layers can be chosen in such a way that the frequencies of the excitonic resonances are close to each other. The ATR spectra calculated for this situation are depicted in Figs. \ref{fig:reflectance-2l-eq}(a),(b). As it is can be seen from the insets, in this case (that can be called double resonance) it is possible to achieve the situation where three reflectivity minima correspond to one value of $\theta$. We also checked another special situation where the higher energy exciton transition in $a_1$ QDs matches the ground state transition in the $a_2$ dots [see Figs. \ref{fig:reflectance-2l}(c), (d)] but apparently no new effects arise in this case and there are only two resonance bands (at $\hbar \omega\approx 2.1\,$eV and $\hbar \omega\approx 2.5\,$eV) seen.

\begin{figure}
\begin{center}
\includegraphics*[width=8cm]{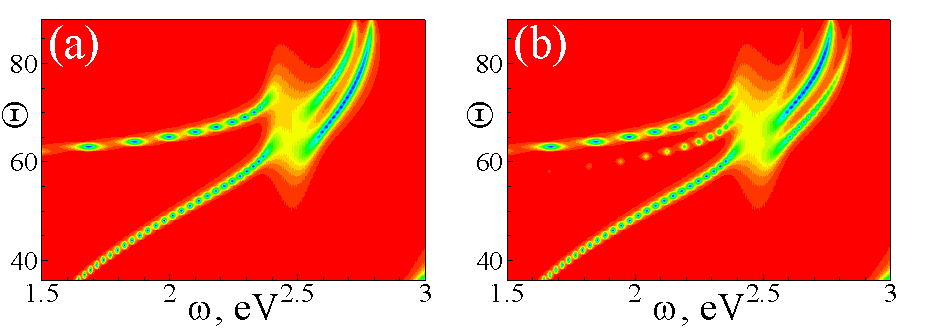}
\end{center}
\caption{(color on line) (Panel a) Reflectivity \textit{versus} angle of
incidence and frequency for the ATR structure consisting of the glass prism, one silver film of thickness $d=53.3\,$nm, one CdSe QD--PMMA composite layer with average QD radius $a_1=3\,$nm and thickness $d_1=300\,$nm, and a semiinfinite silver layer (a) or a second silver film of thickness  $d_{Au}^2=69.23\,$nm and a semi-infinite CdSe QD--PMMA composite with average QD radius $a_1$ (b). }
\label{fig:reflectance-3l}
\end{figure}

Finally, we considered still another potentially interesting sandwich-type structure containing two [Fig. \ref{fig:reflectance-3l}(a)] or three [Fig. \ref{fig:reflectance-3l}(b)] interfaces between silver and CdSe QD--PMMA composite. For such structures, the calculated reflectivity spectra contain two or three minima (for a gived frequency), respectively. One can observe a rather sophisticated "bunching" of modes in the vicinity of the excitonic resonance.

In conclusion, we have shown that the resonance coupling between the plasmon-polaritons propagating along the metal/NC-layer interface and excitons confined in chemically synthesized semiconductor NCs, experimentally observed in Ref. \onlinecite {Gomez10}, can be described theoretically using the appropriate effective dielectric function for the NC composite layer and standard multilayer optics. The SPP-exciton interaction produces a considerable effect on the optical properties of the structure if the dispersion of the NC size in the composite layer is not too large.
We have shown that combining several composite layers with appropriately sized quantum dots and/or more than one metallic films can result in interesting interactions between the various SPP and exciton modes. Owing to these interactions, the energy of an incident electromagnetic wave can be distributed, by means of surface plasmons, between the different QD species, as it has been suggested for molecules adsorbed on a metallic surface \cite {Weiderrecht}. In particular, it opens the possibility to control the relative intensity of light of different colors, emitted by the QDs of different sizes. The work was supported by the FCT (Portugal) under the grant PTDC/FIS/113199/2009.

%-------------------------------------------------------------------------------
% References
%-------------------------------------------------------------------------------


\begin{thebibliography}{99}

\bibitem{Rogach2008}  A. L. Rogach (ed.), {\it Semiconductor Nanocrystal Quantum Dots}, (Springer-Verlag: Wien, 2008).

\bibitem{Basko2000} D. M. Basko, V. M. Agranovich, F. Bassani, ang G. C. La Rocca, Eur. Phys. J. B {\bf 13}, 653 (2000).

\bibitem{Chen2010} Zheyuan Chen†, S. Berciaud, C. Nuckolls, T. F. Heinz, and L. E. Brus, ACS Nano {\bf 4}, 2964 (2010).

\bibitem{Gomez10}  D. E. Gomez, R. C. Vernon, P. Mulvaney and T. J. Davis, Nano Lett. {\bf 10}, 274 (2010).

\bibitem{MG}  J. C. Maxwell--Garnett, {Philos. Trans. R. Soc. London} {\bf 203%
}, 385 (1904).

\bibitem{Bruggeman}  D. A. G. Bruggeman, Ann. Physik (Leipzig) {\bf 24}, 636
(1935).

\bibitem{VA96}  M. I. Vasilevskiy and E. V. Anda, {Phys. Rev. B} {\bf 54}, 5844
(1996).

\bibitem{MV00}  M. I. Vasilevskiy, phys. stat. sol. (b) {\bf 219}, 197 (2000).

\bibitem{Brus} L. E. Brus, J. Chem. Phys. {\bf 80}, 4403 (1984);

\bibitem{c:SPP-theory} J. M. Pitarke, V. M. Silkin, E. V. Chulkov, and P. M. Echenique, Rep. Prog. Phys. {\bf 70}, 1 (2007).

\bibitem{Ag_df}  P. B. Johnson and R. W. Christy, {Phys. Rev. B} {\bf 6}, 4370
(1972).

\bibitem{Weiderrecht} G. P. Weiderrecht, J. E. Hall and A. Boutheiler, {Phys. Rev. Lett.} {\bf 98}, 083001
(2007).

\end{thebibliography}
\end{document}